\documentclass[preprint2]{aastex}

\newcommand{\hm}{H$_2$}
\slugcomment{}

\shorttitle{Hydrogenation of PAH cations.}
\shortauthors{Boschman et al.}

\begin{document}


\title{Hydrogenation of PAH cations: a first step towards H$_2$ formation.}


\author{L. Boschman\altaffilmark{1,2}, G. Reitsma \altaffilmark{2}, S. Cazaux\altaffilmark{1}, T. Schlath\"{o}lter\altaffilmark{2}, R. Hoekstra\altaffilmark{2}, M. Spaans\altaffilmark{1} \and O. Gonz\'{a}lez-Maga\~{n}a\altaffilmark{2}}
%\affil{Kapteyn Astronomical Institute, P.O. Box 800, 9700 AV Groningen, the Netherlands
%\and KVI Atomic and Molecular Physics group, Zernikelaan 25, 9747 AA Groningen, the Netherlands}
\email{boschman@astro.rug.nl}


\altaffiltext{1}{Kapteyn Astronomical Institute, {University of Groningen}, P.O. Box 800, 9700 AV Groningen, the Netherlands}
\altaffiltext{2}{KVI Atomic and Molecular Physics, {University of Groningen}, Zernikelaan 25, 9747 AA Groningen, the Netherlands.}
%\altaffiltext{3}{present address: Center for Astrophysics, 60 Garden Street, Cambridge, MA 02138}
%\altaffiltext{4}{Visiting Programmer, Space Telescope Science Institute}
%\altaffiltext{5}{Patron, Alonso's Bar and Grill}


\begin{abstract}
Molecular hydrogen is the most abundant molecule in the universe.
A large fraction of {H$_{2}$} forms by association of hydrogen atoms adsorbed on polycyclic aromatic hydrocarbons (PAHs), where formation rates depend crucially on the H sticking probability. 
We have experimentally studied PAH hydrogenation by exposing coronene cations, confined in a radiofrequency ion trap, to gas phase atomic hydrogen. 
A systematic increase of the number of H atoms adsorbed on the coronene with the time of exposure is observed. 
Odd coronene hydrogenation states dominate the mass spectrum up to 11 H atoms attached. This indicates the presence of a barrier preventing H attachment to these molecular systems. 
For the second and {fourth} hydrogenation, barrier heights of 72 $\pm$ 6 meV and 40 $\pm$ 10 meV, respectively are found which is in good agreement with theoretical predictions for the hydrogenation of neutral PAHs. 
Our experiments however prove that the barrier does not vanish for higher hydrogenation states. 
These results imply that PAH cations, as their neutral counterparts, exist in highly hydrogenated forms in the interstellar medium. 
Due to {this} catalytic activity, PAH cations and neutrals seem to contribute similarly to the formation of {H$_{2}$}.

\end{abstract}


%astrochemistry; ISM: molecules}%, H$_2$, PAHs}



\section{Introduction}
\label{sec:introduction}
Molecular hydrogen is the most abundant molecule in the universe and the main constituent of regions where stars are forming. \hm\ plays an important role in the chemistry of the interstellar medium, and its formation governs the transformation of atomic diffuse clouds into molecular clouds. Because of the inefficient gas phase routes to form \hm, dust grains have been recognized to be the favored habitat to form \hm\ molecules (\citealt{oort1946}, \citealt{gould1963}). 
The sticking of H atoms onto surfaces has received considerable attention because this mechanism governs the formation of \hm, but also other molecules that contain H atoms. The sticking of H atoms onto dust grains can also be an important mechanism to cool interstellar gas (\citealt{spaans2000}). In the past decades, a plethora of laboratory experiments and theoretical models have been developed to understand how \hm\ forms. As H atoms arrive on dust surfaces, they can be weakly (physisorbed) or strongly (chemisorbed) bound to the surface. The sticking of H in the physisorbed state (\citealt{pirronello1997a}, \citeyear{pirronello1999}, \citeyear{pirronello2000}; \citealt{perry2003}) and in the chemisorbed state (\citealt{zecho2002}; \citealt{hornekaer2006}; \citealt{mennella2006}) has been highlighted by several experiments on different types of surfaces (amorphous carbon, silicates, graphite). 

In the ISM, dust grains are mainly carbonaceous or silicate particles with various sizes and represent an important surface for the formation of \hm. However, a large part ($\sim$ 50$\%$) of the available surface area for chemistry is in the form of very small grains or PAHs (\citealt{Weingartner2001}). These PAHs are predicted to have characteristics similar to graphite surfaces:
However, once the first H atom is chemisorbed on the basal plane, subsequent adsorptions of H atoms in pairs appear to be barrierless for the para dimer and with a reduced barrier for the ortho dimer (\citealt{rougeau2006}). \hm\ can then form by involving a pre-adsorbed H atom in monomer (\citealt{sha2002}; \citealt{morisset2003}; \citeyear{morisset2004b}; \citealt{martinazzo2006chem}) or in a para-dimer configuration (\citealt{bachellerie2007}). However, while these routes represent efficient paths to form \hm, the inefficient sticking of H atoms in monomers constitutes an important obstacle to enter the catalytic regime for \hm\ formation. This results in a very low \hm\ formation efficiency on graphitic/PAH surfaces (\citealt{cazaux2011}).

The hydrogenation on the PAH edges has been identified as an important route to form \hm\ in the ISM (\citealt{bauschlicher1998}; \citealt{hirama2004}; \citealt{lepage2009}; \citealt{mennella2012}; \citealt{thrower2012}). 
Density functional theory calculations have shown that the first hydrogenation of neutral coronene is associated with a barrier ($\sim$60 meV) but that subsequent hydrogenation barriers vanish (\citealt{rauls2008}). Recently, coronene films exposed to H/D atoms at high temperature, were studied by means of IR spectroscopy (\citealt{mennella2012}) and mass spectrometry (\citealt{thrower2012}). 
These measurements showed that  neutral PAHs, when highly hydrogenated, are efficient catalysts for the formation of \hm, and confirmed the high \hm\ formation rate attributed to PAHs in PDRs (\citealt{mennella2012}).

PAH cations, which are usually present at lower extinction A$_{\rm{V}}$, and therefore reside at the surfaces of PDRs, also represent an important route to form \hm\ (\citealt{bauschlicher1998}; \citealt{lepage2009}). 
The addition of the first H atom is predicted to be barrierless. This reaction is exothermic but the product should be stabilized by IR emission. A second H atom can react with the already adsorbed H to form \hm\ without a barrier (\citealt{bauschlicher1998}; \citealt{hirama2004}). 

In this letter, we study experimentally the hydrogenation of coronene cations in the gas phase through exposure to hydrogen atoms. By using mass spectrometry, we show that odd hydrogenation states of coronene cations predominantly populate the mass spectrum. Our results highlight the fact that the further  hydrogenation of PAH cations is associated with a barrier if the number already attached H atoms is odd, and no barrier if this number is even. This alternanting barrier-no barrier occurence seems to remain with increasing hydrogenation. These results  suggest that PAH cations can also enjoy highly hydrogenated states in the interstellar medium, and acts as catalysts for \hm\ formation.

\section{Experiments}
\label{sec:experiments}
In this pilot experiment we show the feasibility of studying the hydrogenation of PAHs in the gas phase. For this purpose, {we use a setup designed to study molecular ions in a radiofrequency ion trap.}
Time-of-flight mass spectrometry of the trap content is used to identify the changes in mass of the coronene cations and therefore deduce their respective degrees of hydrogenation.

\subsection{Set-up}
\label{ssec:setup}
\begin{figure}
  \resizebox{0.5\textwidth}{!}{\includegraphics{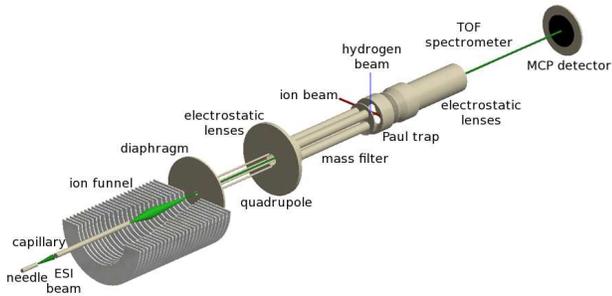}}
  \caption{The setup used, with the ion funnel, quadrupoles, ion trap, hydrogen source and detector}
  \label{fig:setup}
\end{figure}

The experiments have been performed using a home-built tandem-mass spectrometer shown schematically in  figure \ref{fig:setup} (\citealt{bari2011}).  A beam of singly charged coronene radical cations ([C$_{24}$H$_{12}$]$^+$, m/z 300) was extracted from an electrospray ion source. The ions were phase-space compressed in an radiofrequency (RF) ion funnel and subsequently in an RF quadrupole ion guide. Mass selection was accomplished by using an RF quadrupole mass filter. Accumulation of the ions took place in a three dimensional RF ion trap (Paul trap). A He buffer gas at room temperature was used to collisionally cool the trapped cations. Exposure to gas-phase atomic hydrogen for variable periods of time led to multiple hydrogen adsorption on the coronene cations. An electric extraction field was then applied between the trap end-caps to extract the trapped hydrogenated cations into a time-of-flight (TOF) mass spectrometer with resolution M/$\Delta$M $\sim$ 200. To obtain mass spectra of sufficient statistics,
 typically a couple of hundred TOF traces were accumulated. 

Electrospray ionization allows to gently transfer ions from the {liquid} phase into the gas phase. Inspired by the method of \cite{maziarz2005} we have run the ion source with a solution consisting of 600 $\mu$L of saturated solution of coronene in methanol, 350 $\mu$L of HPLC grade methanol and 50 $\mu$L of 10 mM solution of $\mathrm{AgNO_{3}}$ solution in methanol. In the liquid phase, electron transfer from a coronene molecule to a silver ion leads to formation of the required radical cation. 

The trapped ions are exposed to hydrogen atoms produced from \hm\ by a Slevin type source which has been extensively used  in crossed beam experiments (\citealt{hoekstra1991},\citealt{bliek1998}). While in the earlier work the dissociation fractions were determined by means of electron impact excitation or HeII line emission, we now use charge removal (captured ionization) and dissociation induced by 40 keV {He}$^{2+}$. For these processes the cross sections are well-known (\citealt{shah1978}). In this way we determine a hydrogen dissociation fraction of $n \left( \mathrm{H} \right) / \left( n \left( \mathrm{H} \right) + n \left( \mathrm{H_{2}} \right) \right) \approx 0.3$. The temperature of the H beam is around room temperature ($\sim$25~meV).

\subsection{Results}
\label{ssec:experiments}

Coronene ions are exposed to a constant flux of H atoms for different periods of time before their degree of hydrogenation is determined by means {of} mass spectrometry. The irradiation time is varied from 1.0 up to 30 s to study the time-dependence of coronene hydrogenation.

\begin{figure*}
  \resizebox{0.33\textwidth}{!}{\includegraphics{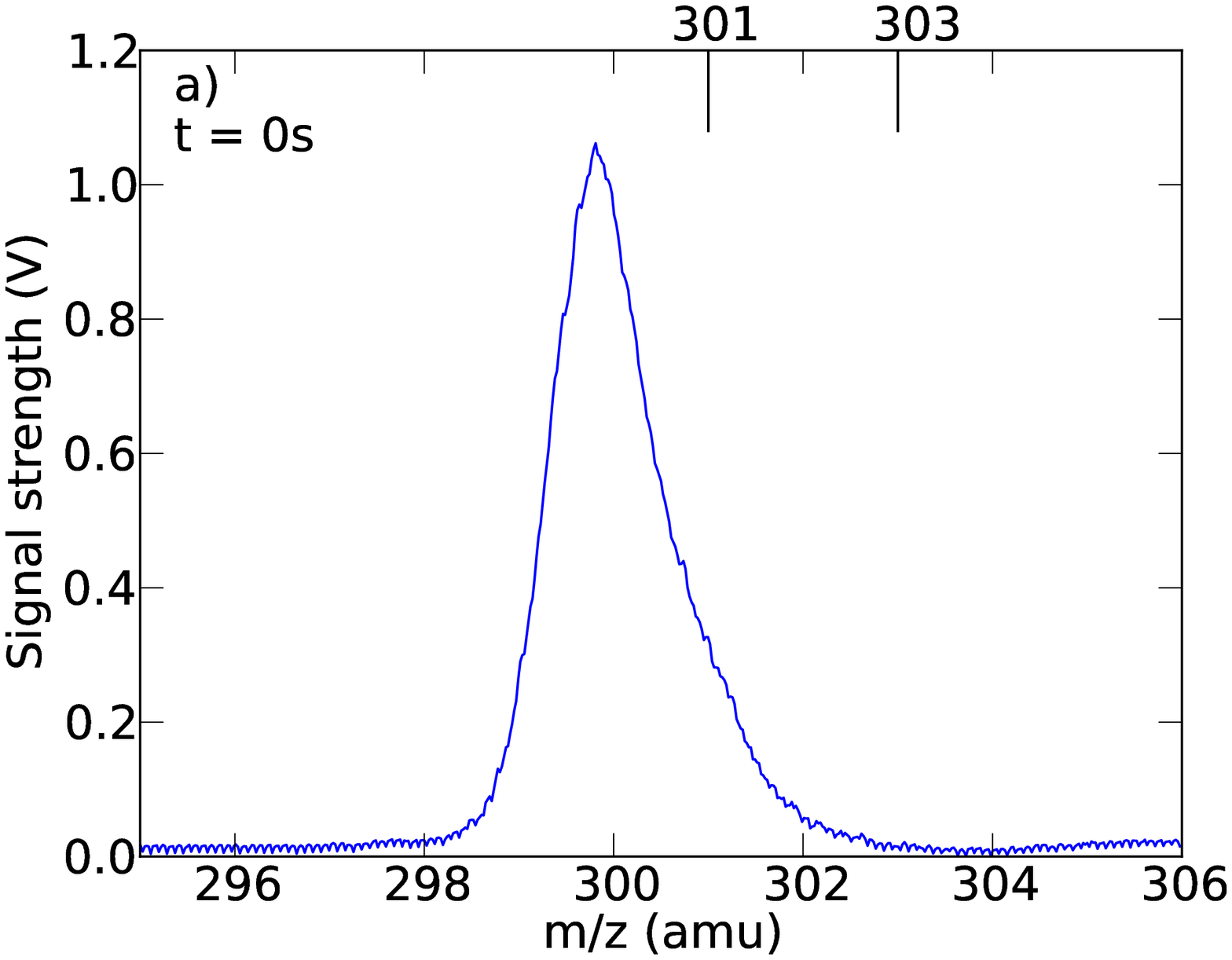}}
  \resizebox{0.33\textwidth}{!}{\includegraphics{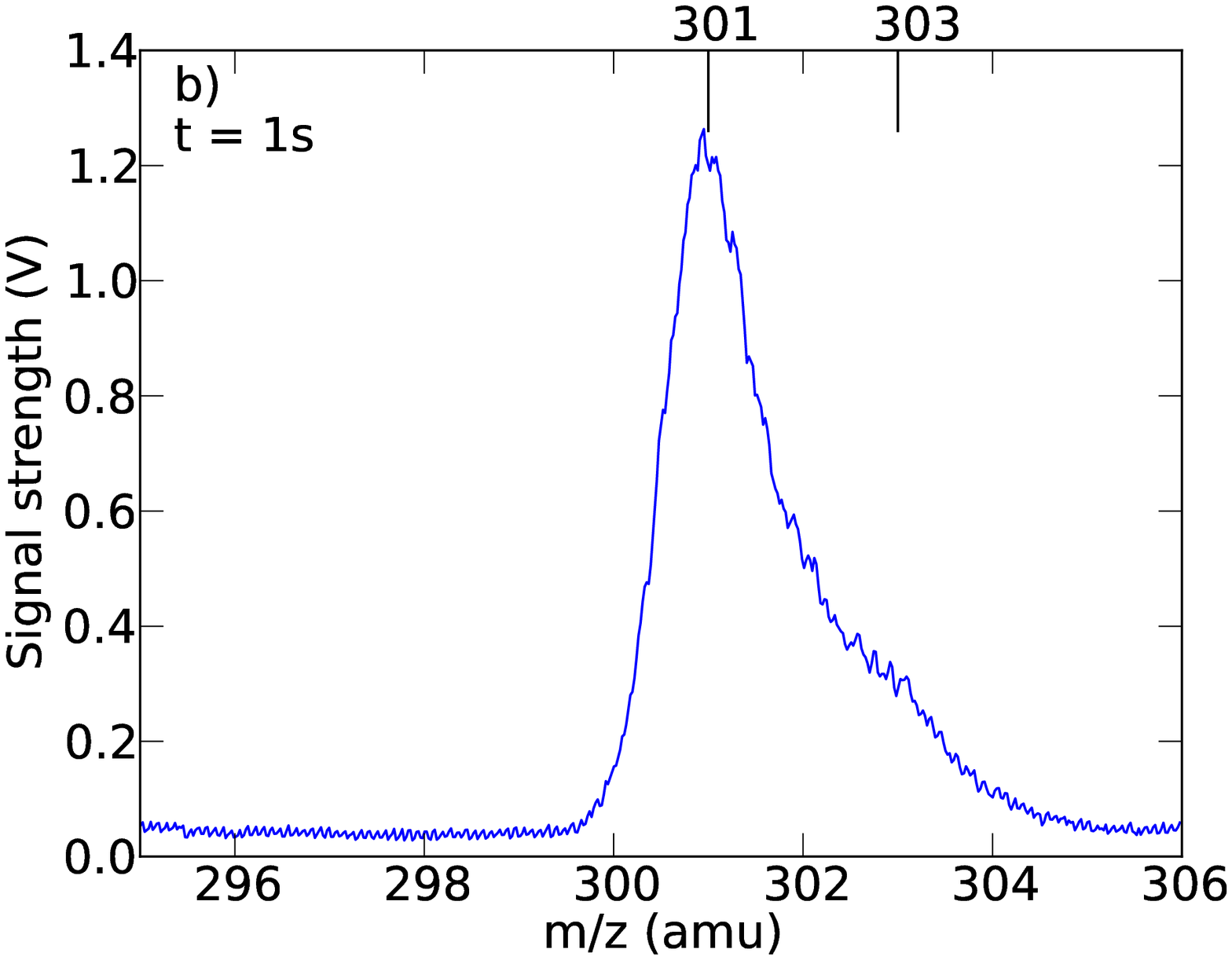}}
%  \resizebox{0.33\textwidth}{!}{\includegraphics{072_coronene_cation_T261_r01_50.eps}}
  \resizebox{0.33\textwidth}{!}{\includegraphics{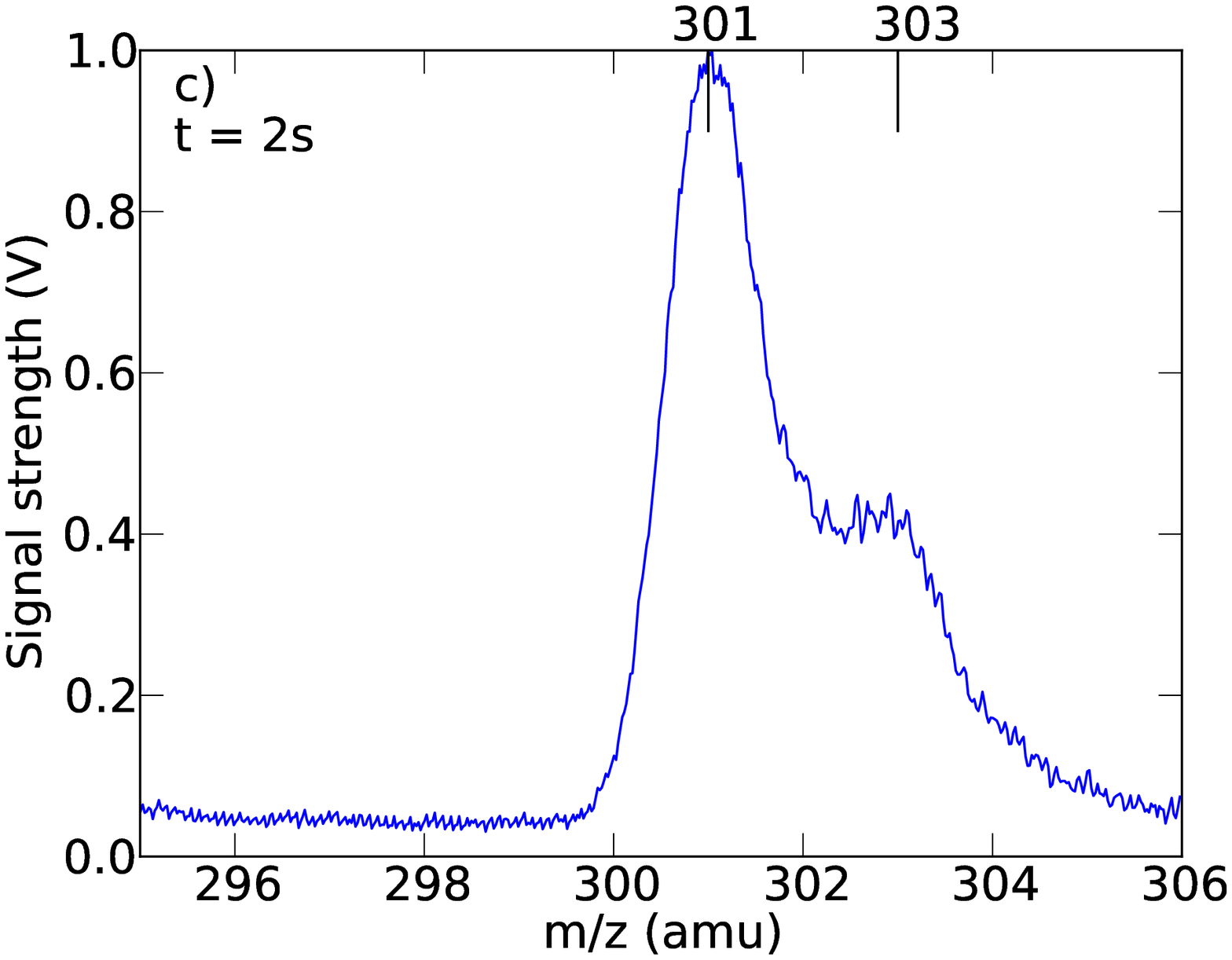}}
%  \resizebox{0.33\textwidth}{!}{\includegraphics{073_coronene_cation_T261_r02_50.eps}}
  \resizebox{0.33\textwidth}{!}{\includegraphics{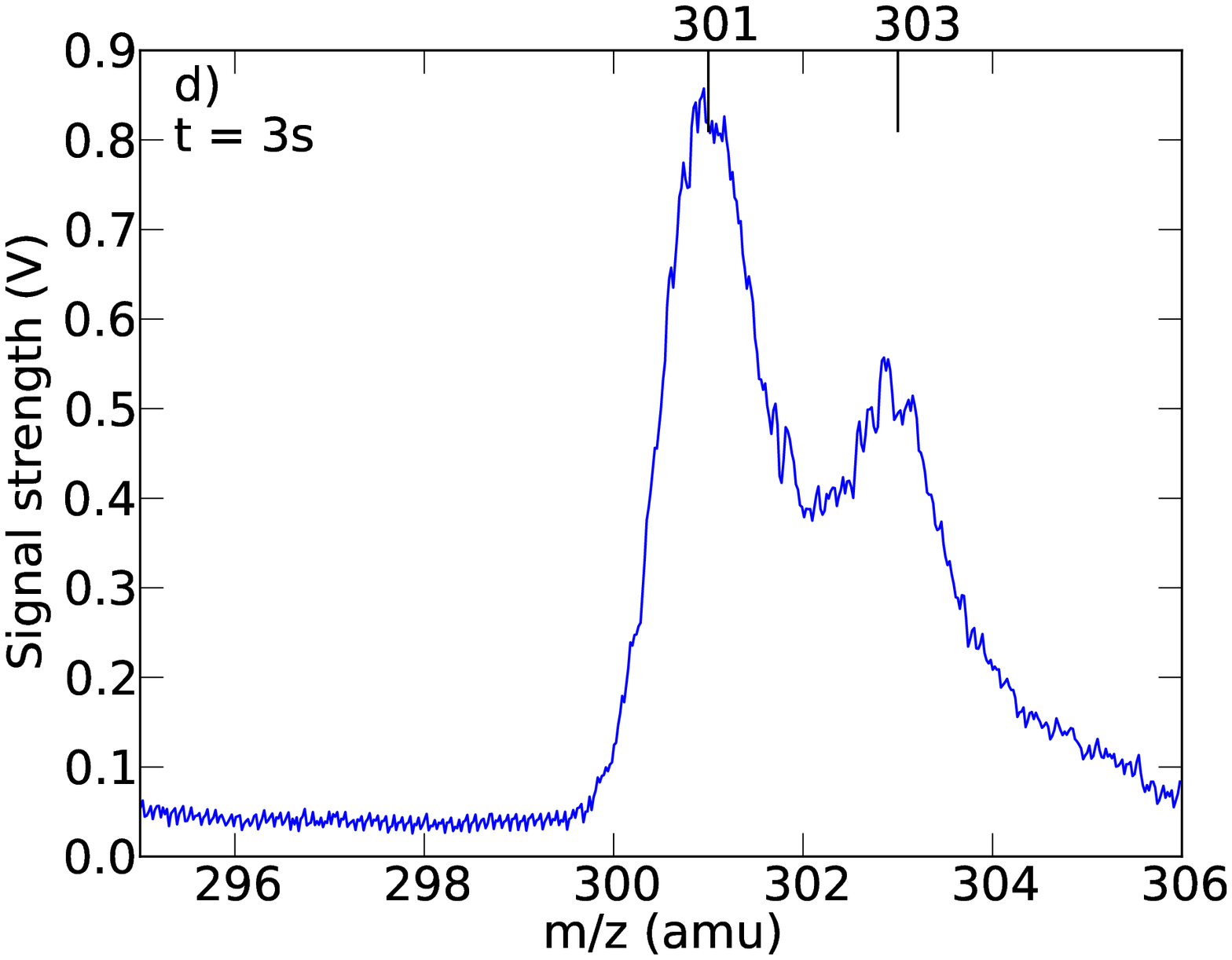}}
%  \resizebox{0.33\textwidth}{!}{\includegraphics{075_coronene_cation_T261_r03_50.eps}}
  \resizebox{0.33\textwidth}{!}{\includegraphics{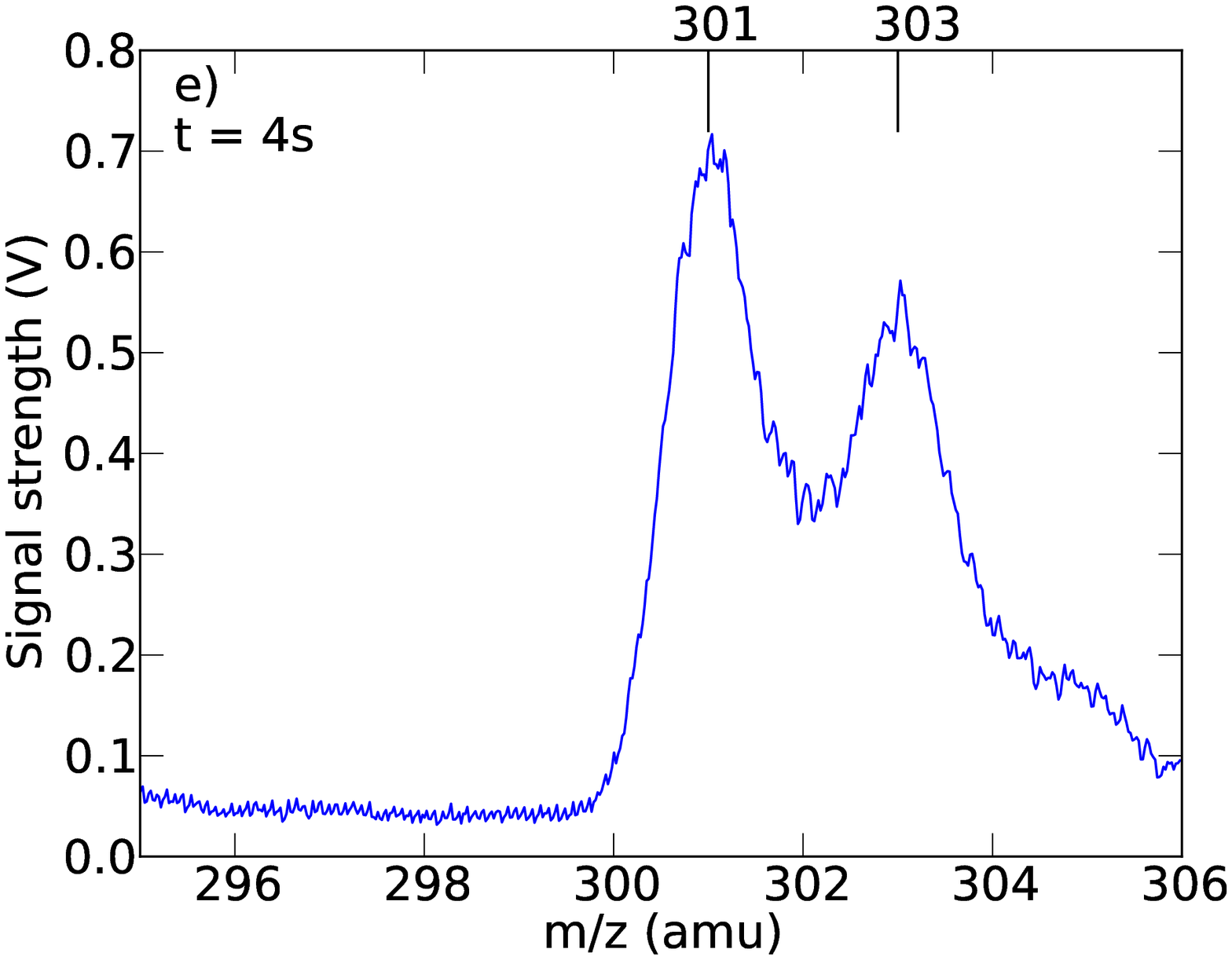}}
%  \resizebox{0.33\textwidth}{!}{\includegraphics{074_coronene_cation_T261_r04_50.eps}}
  \resizebox{0.33\textwidth}{!}{\includegraphics{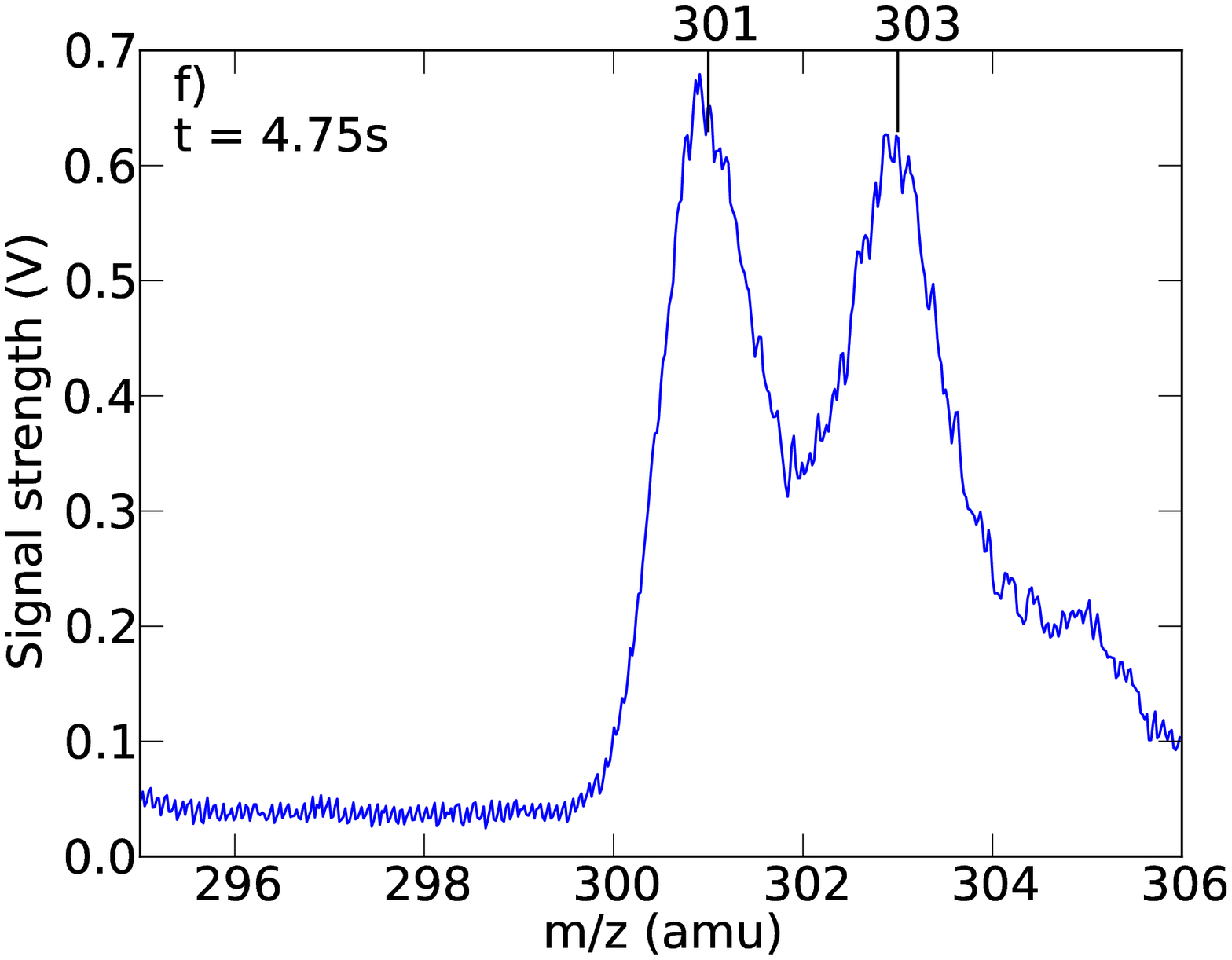}}
  \caption{Mass spectrum of coronene {a) without and with exposure to} H atoms during b) 1 s c) 2 s d) 3 s e) 4 s f) 4.5 s.}
  \label{fig:rawdata}
\end{figure*}

The data obtained from our experiment are a series of mass spectra of hydrogenated coronene cations as a function of H exposure time. Some of the spectra are shown in fig.\ref{fig:rawdata}. Fig.\ref{fig:rawdata}(a) shows the mass spectrum of the native m/z=300 coronene cations. {A similar, thus unchanged, mass spectrum is obtained (not shown in this article) if we irradiate coronene cations with molecular hydrogen. This means that molecular hydrogen does not stick to coronene cations at room temperature.}

\begin{figure*}
  \resizebox{0.4\textwidth}{!}{\includegraphics{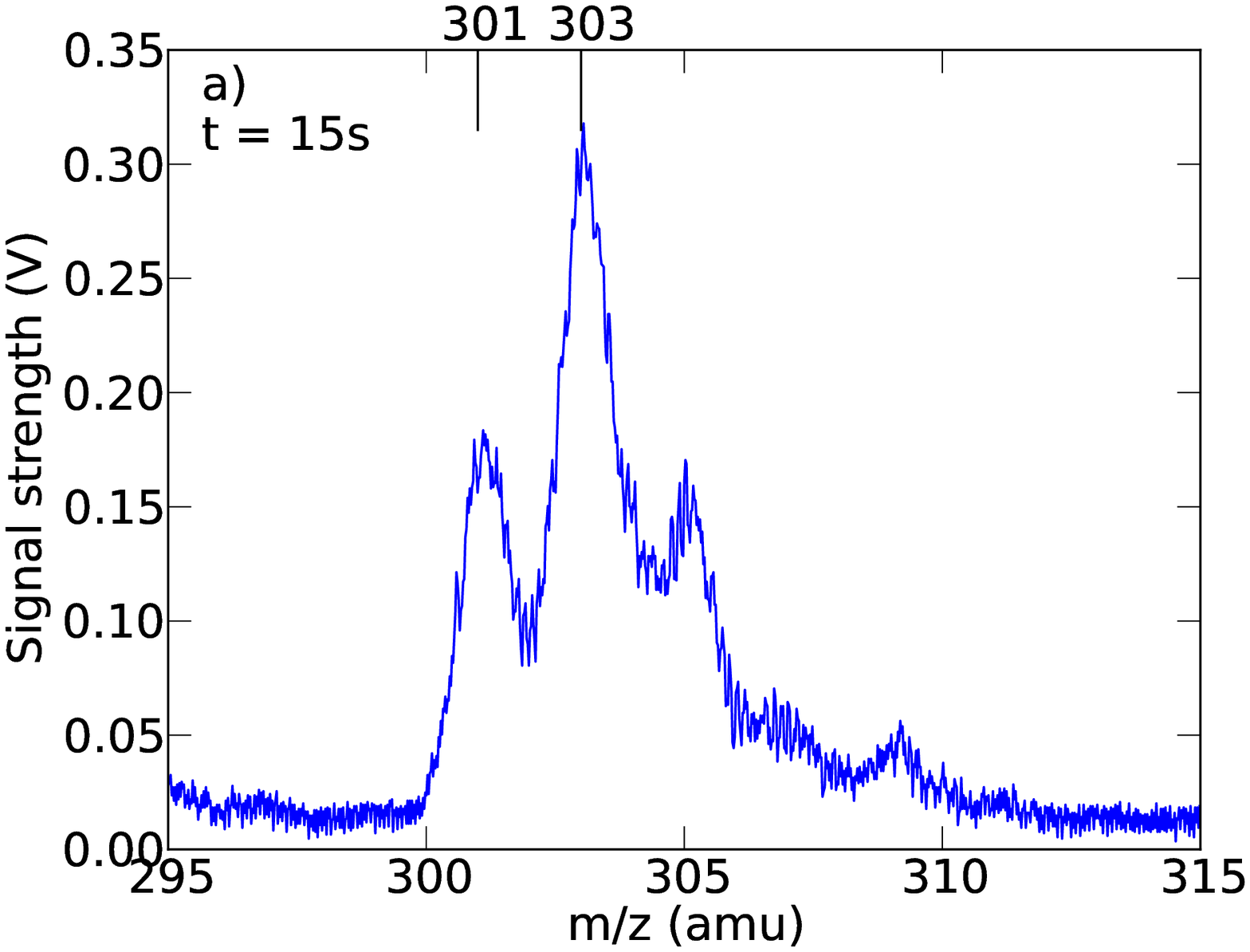}}
  \resizebox{0.4\textwidth}{!}{\includegraphics{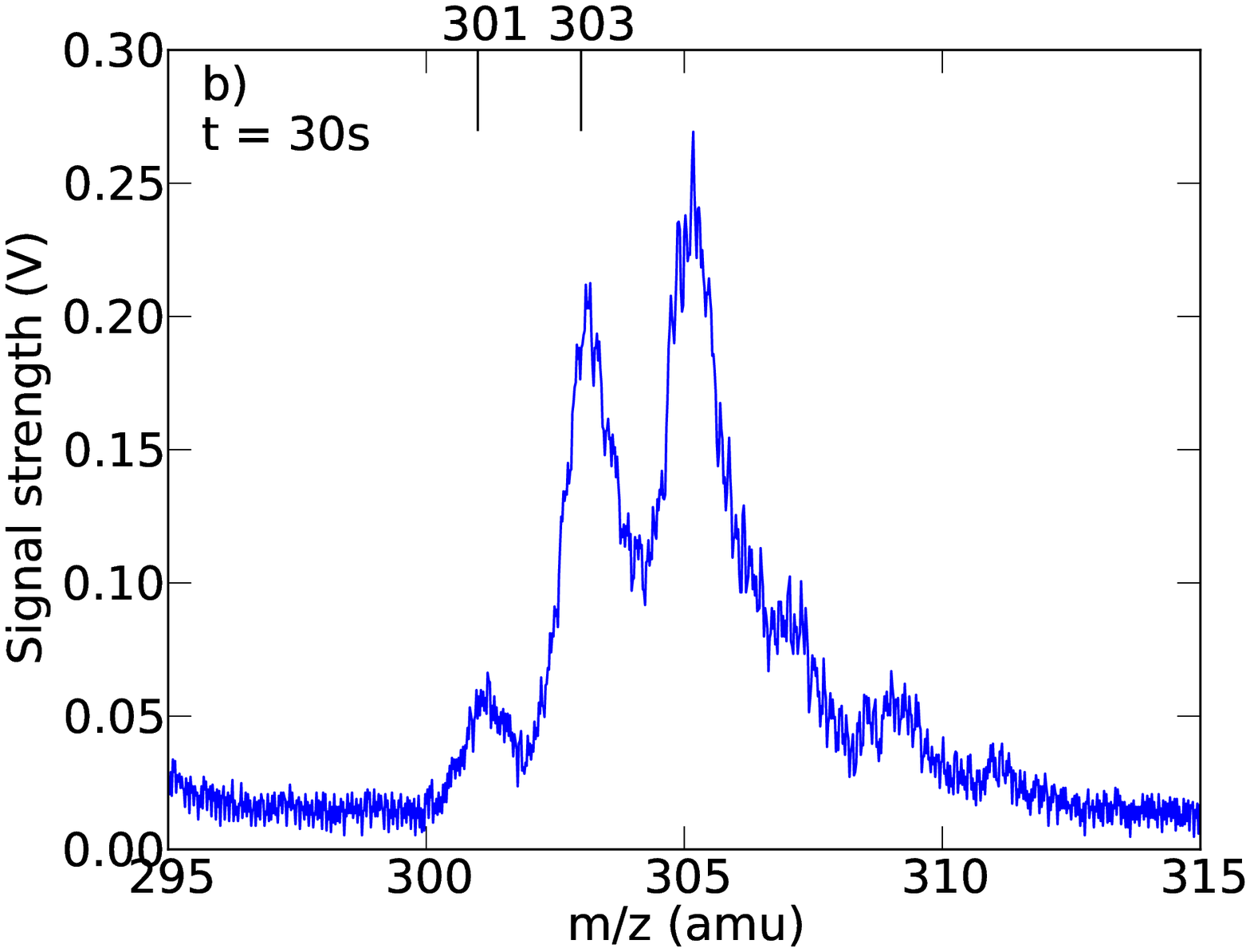}}
  \caption{Same as fig.\ref{fig:rawdata}  for much longer H exposures of a) 15 s b) 30 s.}
  \label{fig:longdata}
\end{figure*}

After turning on the hydrogen source and exposing the coronene cations to the {atomic} hydrogen beam for 1.0 s (fig.\ref{fig:rawdata}, (b)), the peak at $m/z = 300$ shifts to 301, which means that the trap content main constituent is (C$_{24}$H$_{12}$+H)$^+$. For increasing irradiation time (fig.\ref{fig:rawdata}(c) t= 2 s, (d) 3 s, (e) 4 s and (f) 4.75 s), the peak at $m/z$=301 disappears progressively while a peak at $m/z = 303$ and then  at $m/z = 305$ (for t = 4.75 s see fig.\ref{fig:rawdata}(f) ) appears, which indicates the addition of 3  and 5 hydrogen atoms, respectively. At longer exposure time  (fig.\ref{fig:longdata}(a) t $\sim$15 s), the $m/z$=303 peak dominates the signal, and a peak at $m/z$=305 appears. At even longer irradiation times  (fig.\ref{fig:longdata}(b) t $\sim$30~s),  the peak $m/z$=305 dominates and peaks at $m/z$=307 and 309 appear.
These peaks clearly show the evolution of the hydrogenation states of coronene cations with H irradiation time. 

\section{Analysis and discussion}
Our results show that the most important peaks measured in the mass spectrum shift from lower masses to higher masses with increasing H exposure time. 
In order to follow the evolution of the first hydrogenated state of coronene cation (C$_{24}$H$_{12}$+H)$^+$ (CorH$^+$) to the second  (C$_{24}$H$_{12}$+2H)$^+$ (CorH$_2^+$), third (CorH$_3^+$) and {fourth} (CorH$_4^+$) hydrogenated states, we use a simple model that describes this evolution :

\begin{equation}
\frac{d ~ n_{\mathrm{CorH^{+}}} }{dt} = - \left(A_{2} ~ e^{-\frac{E_{2}}{k_{B} T_{\mathrm{{gas}}}}} n_{\mathrm{CorH^{+}}}\right) n_{H}
\end{equation}
\begin{equation}
\frac{d ~ n_{\mathrm{CorH_{2}^{+}}} }{dt} = \left( A_{2} ~ e^{-\frac{E_{2}}{k_{B} T_{\mathrm{{gas}}}}} n_{\mathrm{CorH^{+}}} - A_{3} ~n_{\mathrm{CorH_{2}^{+}}} \right) n_{H}
\end{equation}
\begin{equation}
\frac{d ~ n_{\mathrm{CorH_{3}^{+}}} }{dt} = \left( A_{3} ~n_{\mathrm{CorH_{2}^{+}}}-A_{4} ~ e^{-\frac{E_{4}}{k_{B} T_{\mathrm{{gas}}}}} n_{\mathrm{CorH_3^{+}}} \right) n_{H}
\end{equation}
\begin{equation}
\frac{d ~ n_{\mathrm{CorH_{4}^{+}}} }{dt} = \left( A_{4} ~ e^{-\frac{E_{4}}{k_{B} T_{\mathrm{{gas}}}}} n_{\mathrm{CorH_3^{+}}} - A_{5} ~n_{\mathrm{CorH_{4}^{+}}} \right) n_{H}
\end{equation}

Hydrogenation of CorH$_{2n+1}$$^+$ follows an Arrhenius expression where A$_{2n+2}$ is the prefactor and E$_{2n+2}$ is the barrier, while hydrogenation of CorH$_{2n}^+$ follows the same expression with a prefactor A$_{2n+1}$ and no barrier. k$_{B}$ is the Boltzmann constant and T the temperature of the H beam (T$\sim$25~meV).

In these equations we {do not include} abstraction, meaning that the time evolution of the contribution of each state is governed entirely by hydrogenation. 
{This assumption is made in order to derive the first barriers of hydrogenation. Abstraction can be neglected in the conditions of our experiments for low exposure times. 
This is supported by previous experiments where the cross section for addition of hydrogen to neutral coronene} is predicted to be 20 times that for abstraction (\citealt{mennella2012}). {Further support is drawn from a kinetic chemical model we developed, which shows that abstraction has to be very low compared to hydrogenation to be able to mimic the experimental results (Boschman et al. in prep).
However, for long H exposure time we expect the hydrogenation degree of the coronene cations to reach a steady state which will allow us to derive the contribution of abstraction relative to addition, and therefore derive the H$_{2}$ formation rate due to PAH cations. It should also be kept in mind that in the conditions of our experiments, the H atoms are at room temperature meaning that they cross the barriers for abstraction (10 meV, \citealt{rauls2008}) and addition (40 - 60 meV, \citealt{rauls2008}) with similar ease. Under interstellar conditions, however, the abstraction will dominate by 8 orders of magnitude (at 20 K) because of the barrier differences.
}

The first hydrogenation is expected to take place at the outer edge carbon atom (\citealt{hirama2004}). This state provides more conformational freedom to the four neighbouring outer edge carbon atoms, ensuring a preference for the second hydrogenation to take place at one of those four carbon atoms. The third hydrogenation will preferentially take place at the outer edge carbon next to the second H atom. Again, the forth H atom can be bound to one of the four neighbouring outer edge carbon atoms, and the fifth sticks on the neighboring outer edge carbon. This scenario of H atoms sticking preferentially on outer edge carbons next to already adsorbed atoms is described in \cite{rauls2008}.  

\begin{figure*}
  \resizebox{0.33\textwidth}{!}{\includegraphics{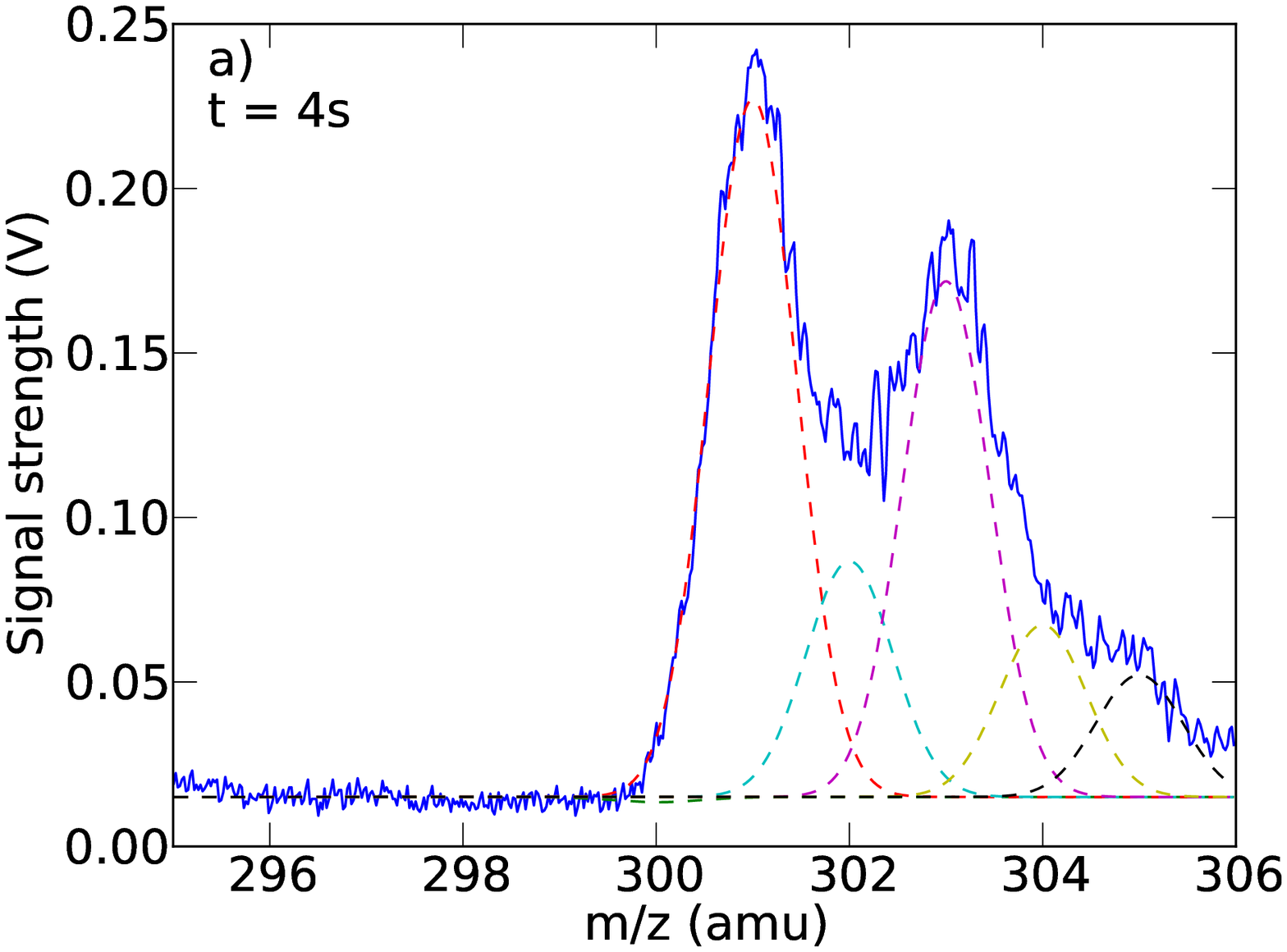}}
  \resizebox{0.33\textwidth}{!}{\includegraphics{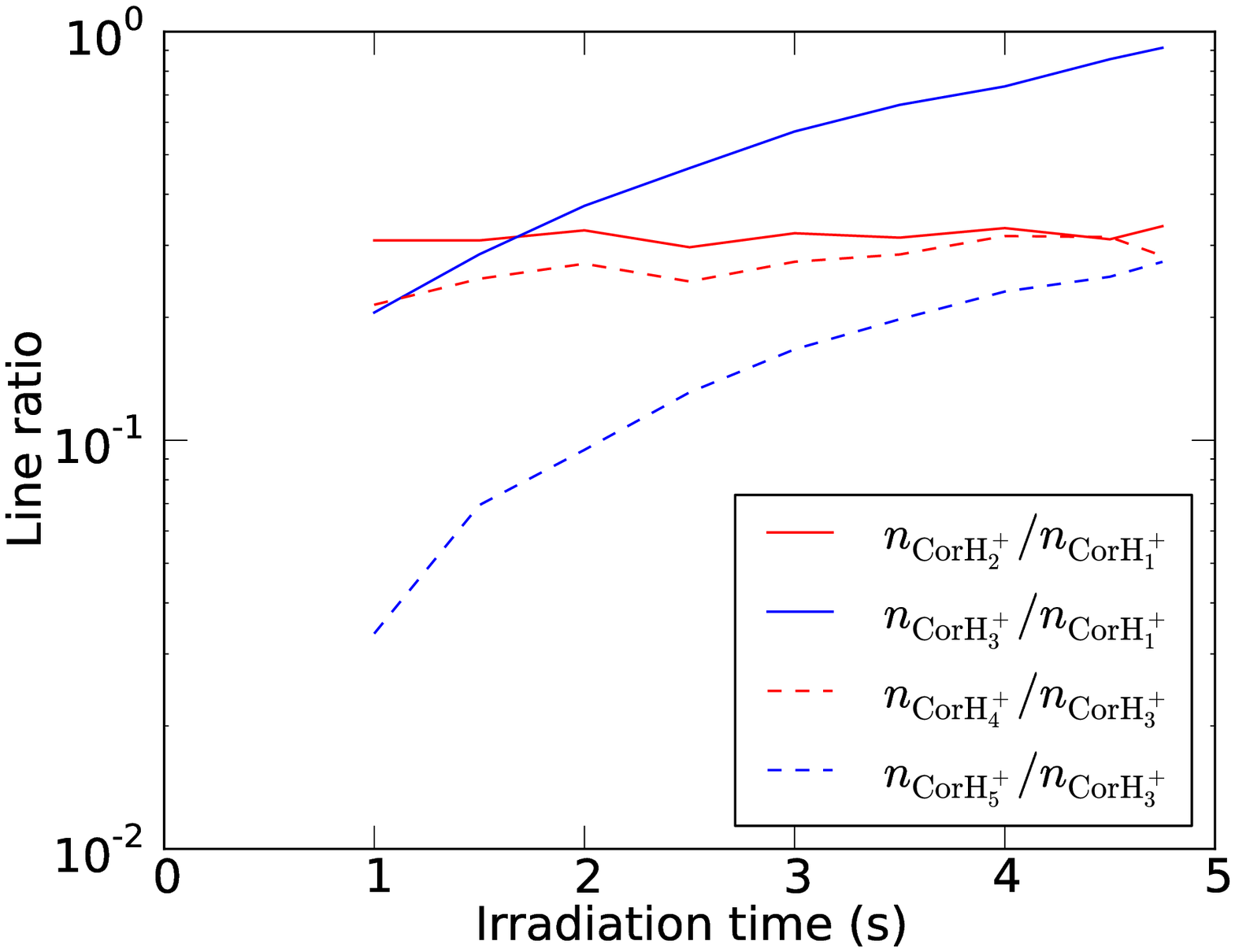}}
  \resizebox{0.33\textwidth}{!}{\includegraphics{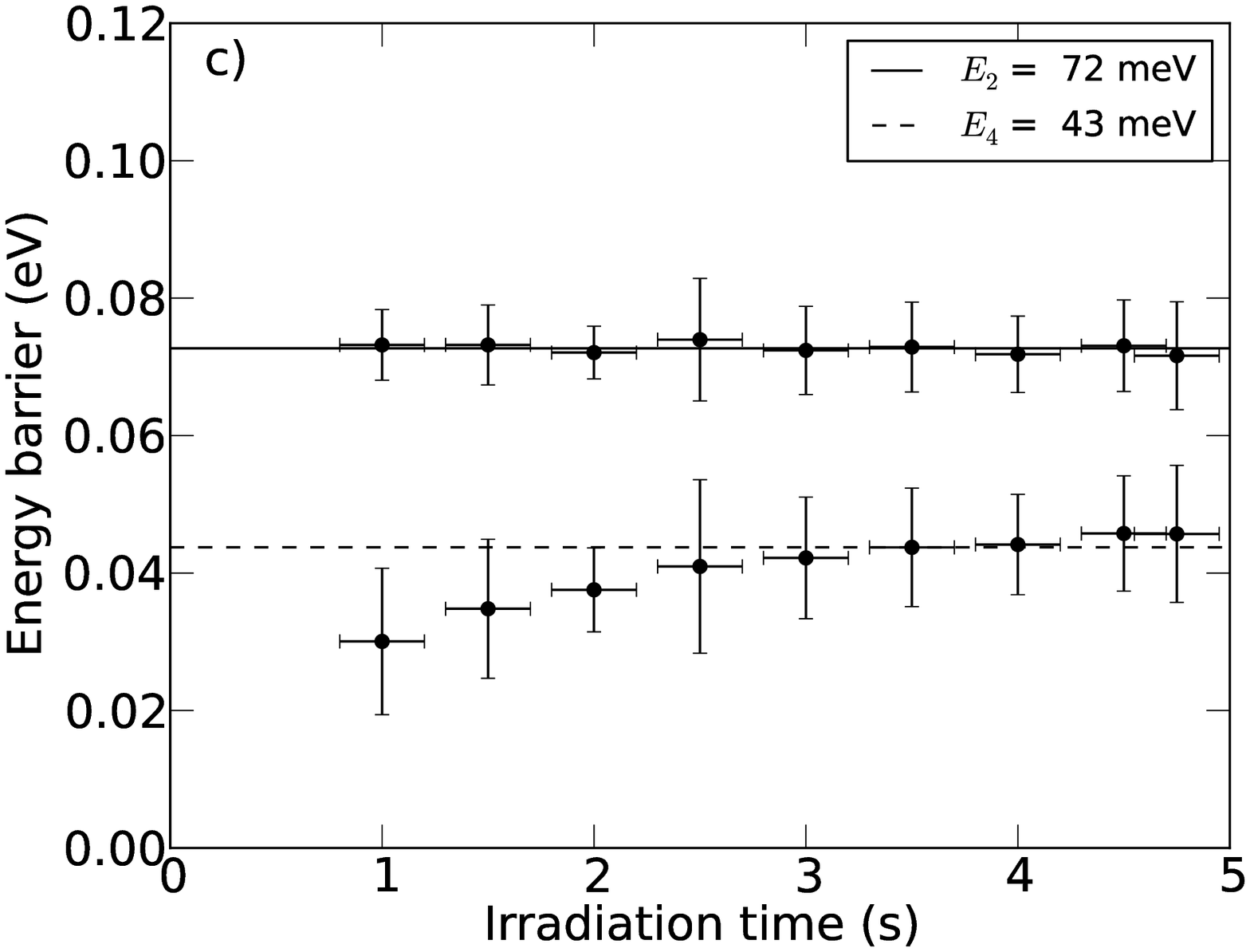}}
  \caption{a) Contribution of every peak determined by fitting our data with Gaussians with identical widths. b) Ratios between different hydrogenation states as function of time. c) Barrier heights for the second and fourth hydrogenations. }
\label{fig:fit}
\end{figure*}

The contribution of every peak is determined by fitting our data with Gaussians with identical widths (see fig.\ref{fig:fit}(a)). The ratios between different hydrogenation states as function of time are reported in fig.\ref{fig:fit}(b). It appears that  the ratio between the contribution of the first (CorH$^+$) and the second (CorH$_2^+$) hydrogenation state does not evolve with time for short time scales $\left( \frac{n_{\mathrm{CorH^{+}}}}{ n_{\mathrm{CorH_{2}^{+}}}} \sim 3 ~\rm{until}\ 5s \right)$. Also, the ratio between the third (CorH$_3^+$) and the forth (CorH$_4^+$) hydrogenation state shows identical behaviour after t$\ge$ 2s  $\left( \frac{n_{\mathrm{CorH_3^{+}}}}{ n_{\mathrm{CorH_{4}^{+}}}} \sim 3 ~\rm{from}\ 2s\ onwards \right)$. Before this exposure time the n$_{\mathrm{CorH_3^{+}}}$ and n$_{\mathrm{CorH_4^{+}}}$ signals are very weak, and the ratio is uncertain.
We can therefore assume that for these measurements $\frac{d}{dt} \left( \frac{ n_{\mathrm{CorH_{2}^{+}}} }{ n_{\mathrm{CorH^{+}}} } \right)=0$ and $\frac{d}{dt}\left( \frac{ n_{\mathrm{CorH_{3}^{+}}} }{ n_{\mathrm{CorH_3^{+}}} } \right)=0$. The expression for the CorH$^{+}$ to CorH$_{2}^{+}$ as well as for the CorH$_3^{+}$ to CorH$_{4}^{+}$ energy barriers can then be written as:

\begin{equation}
  \label{eq:barrier2}
E_2 = - k_{B} T_{\mathrm{{gas}}} \ln \left( \frac{A_{3}}{ A_{2}} \frac{1}{1 + \frac{ n_{\mathrm{CorH^{+}}} }{n_{\mathrm{CorH_{2}^{+}}} } } \right) 
\end{equation}
\begin{equation}
  \label{eq:barrier4}
E_4 = - k_{B} T_{\mathrm{{gas}}} \ln \left( \frac{A_{5}+A_{3}\frac{ n_{\mathrm{CorH_2^{+}}} }{n_{\mathrm{CorH_3}^{+}}}}{A_{4}} \frac{1}{1 + \frac{ n_{\mathrm{CorH_3^{+}}} }{n_{\mathrm{CorH_4}^{+}}} } \right) 
\end{equation}

From these expressions we derive the energy barrier E$_2$ as 72$\pm$6 meV and  E$_4$ as 43$\pm$8 meV, as shown in fig.\ref{fig:fit}(c). This shows that hydrogenation barriers are decreasing with increasing hydrogenation. However, our results also show that odd hydrogenated states dominate the mass spectrum even for high degrees of hydrogenation (fig.\ref{fig:longdata}). This highlights the presence of a barrier-no barrier alternation  from one hydrogenated state to another, up to high hydrogenation states. So our results indicate that even if the hydrogenation barriers decrease for the first hydrogenations, they do not vanish completely and remain at higher hydrogenation states. The barriers derived in our study are similar to the one calculated by \cite{rauls2008} for neutral coronene. This means that the first hydrogenations of coronene cations should be comparable to the hydrogenation of neutral coronene. However, for higher degree of hydrogenation we show that these barriers still exist, while the 
calculations from \cite{rauls2008} predict that these barriers vanish after a few hydrogenations. Recent mass spectrometric measurements of coronene films exposed to H/D atoms do not show preferences for even or odd hydrogenation states of neutral coronene (\citealt{thrower2012}). However, these measurements are not very sensitive to barrier heights well bellow 100 meV, since the experiments were performed with atoms at beam temperature of 170 meV. 

In PDRs exposed to UV fields less than few hundreds G$_0$, the spatial distribution of \hm\ and PAHs does correlate (\citealt{habart2003}, \citealt{habart2005}, \citealt{compiegne2007}) contrary to what is seen in the presence of strong UV fields (\citealt{tielens1993}, \citealt{berne2009}). The \hm\ formation rates have been derived for several PDRs exposed to various UV radiation fields. These rates can be explained by the contribution of PAHs to the formation of \hm\ (\citealt{habart2004}). Depending on the UV intensity, the PAHs observed can either be PAH cations, that are present in regions at low visual exctinctions A$_{\rm{V}}$, or neutral PAHs, which are located at higher extinctions. 
Work by \cite{wolfire2008} and \cite{spaans2005} has shown that high-UV and high density PDRs (n$_{\rm{H}}$$\ge$10$^3$ cm$^{-3}$ and G$_0$$\ge$100 {, G${_{0} = 1.6 \times 10^{-3} \mathrm{{\ erg\ cm^{-3}\ s^{-1}}}}$}) can maintain a $\sim$ 30$\%$ cationic fraction upto a few mag in A$_{\rm{V}}$. More relevant to this work, \cite{cox2006} have studied low-UV PDRs (G$_0$$\le$100), and followed the PAH charge balance for different densities, UV radiation fields and metallicities. They found that PAH cations dominate over neutrals and anions for A$_{\rm{V}}$$\le$2 mag. The \hm\ formation rates observed in PDRs exposed to different UV fields can therefore be partly attributed to neutral and cationic PAHs.

Our results show that the hydrogenation processes of neutral and cationic PAHs are similar and should contribute similarly to the formation of \hm. Further experimental investigations will allow us to derive the \hm\ formation rate for PAH cations.

\section{Conclusions}
We have investigated the addition of hydrogen atoms to coronene cations in the gas phase and observed increasing hydrogenation with H exposure time. Our results show that odd hydrogenated states dominate the mass spectrum, which evidences the presence of a barrier for the further hydrogenation of odd hydrogenation states. The first hydrogen sticks to the coronene cations without a barrier (\citealt{snow1998}, \citealt{hirama2004}). The second and forth hydrogenations are associated with barriers of about 72 $\pm$ 6 meV and 43 $\pm$ 8 meV, while the third and fifth hydrogenation are barrierless. These barriers are similar to the one calculated for neutral coronene (\citealt{rauls2008}). Our results indicate that superhydrogenated PAH cations {(\cite{li2012})} should also be found in the interstellar medium, and be important catalysts for the formation of \hm, as it is the case for their neutral counterparts.

\subsection*{ACKNOWLEDGMENTS}
 L. B. and S. C. are supported by the Netherlands Organization for Scientific Research (NWO). G.R. recognizes the funding by the NWO Dutch Astrochemistry Network.
We would like to thank the anonymous referee for the helpful comments.


\begin{thebibliography}{39}
\expandafter\ifx\csname natexlab\endcsname\relax\def\natexlab#1{#1}\fi

\bibitem[{{Bachellerie} {et~al.}(2007){Bachellerie}, {Sizun}, {Teillet-Billy},
  {Rougeau}, \& {Sidis}}]{bachellerie2007}
{Bachellerie}, D., {Sizun}, M., {Teillet-Billy}, D., {Rougeau}, N., \& {Sidis},
  V. 2007, Chemical Physics Letters, 448, 223

\bibitem[{{Bari} {et~al.}(2011){Bari}, {Gonzalez-Magana}, {Reitsma}, {Werner},
  {Schippers}, {Hoekstra}, \& {Schlatholter}}]{bari2011}
{Bari}, S., {Gonzalez-Magana}, O., {Reitsma}, G., {et~al.} 2011, The Journal of
  Chemical Physics, 134, 024314

\bibitem[{{Bauschlicher}(1998)}]{bauschlicher1998}
{Bauschlicher}, Jr., C.~W. 1998, \apjl, 509, L125

\bibitem[{{Bern{\'e}} {et~al.}(2009){Bern{\'e}}, {Fuente}, {Goicoechea},
  {Pilleri}, {Gonz{\'a}lez-Garc{\'{\i}}a}, \& {Joblin}}]{berne2009}
{Bern{\'e}}, O., {Fuente}, A., {Goicoechea}, J.~R., {et~al.} 2009, \apjl, 706,
  L160

\bibitem[{Bliek {et~al.}(1997)Bliek, Woestenenk, Hoekstra, \&
  Morgenstern}]{bliek1998}
Bliek, F., Woestenenk, G., Hoekstra, R., \& Morgenstern, R. 1997, Hyperfine
  Interactions, 108, 121

\bibitem[{{Cazaux} {et~al.}(2011){Cazaux}, {Morisset}, {Spaans}, \&
  {Allouche}}]{cazaux2011}
{Cazaux}, S., {Morisset}, S., {Spaans}, M., \& {Allouche}, A. 2011, \aap, 535,
  A27

\bibitem[{{Compi{\`e}gne} {et~al.}(2007){Compi{\`e}gne}, {Abergel},
  {Verstraete}, {Reach}, {Habart}, {Smith}, {Boulanger}, \&
  {Joblin}}]{compiegne2007}
{Compi{\`e}gne}, M., {Abergel}, A., {Verstraete}, L., {et~al.} 2007, \aap, 471,
  205

\bibitem[{{Cox} \& {Spaans}(2006)}]{cox2006}
{Cox}, N.~L.~J. \& {Spaans}, M. 2006, \aap, 451, 973

\bibitem[{{Gould} \& {Salpeter}(1963)}]{gould1963}
{Gould}, R.~J. \& {Salpeter}, E.~E. 1963, \apj, 138, 393

\bibitem[{{Habart} {et~al.}(2005){Habart}, {Abergel}, {Walmsley}, {Teyssier},
  \& {Pety}}]{habart2005}
{Habart}, E., {Abergel}, A., {Walmsley}, C.~M., {Teyssier}, D., \& {Pety}, J.
  2005, \aap, 437, 177

\bibitem[{{Habart} {et~al.}(2003){Habart}, {Boulanger}, {Verstraete}, {Pineau
  des For{\^e}ts}, {Falgarone}, \& {Abergel}}]{habart2003}
{Habart}, E., {Boulanger}, F., {Verstraete}, L., {et~al.} 2003, \aap, 397, 623

\bibitem[{{Habart} {et~al.}(2004){Habart}, {Boulanger}, {Verstraete},
  {Walmsley}, \& {Pineau des For{\^e}ts}}]{habart2004}
{Habart}, E., {Boulanger}, F., {Verstraete}, L., {Walmsley}, C.~M., \& {Pineau
  des For{\^e}ts}, G. 2004, \aap, 414, 531

\bibitem[{Hirama {et~al.}(2004)Hirama, Tokosumi, Ishida, \& ichi
  Aihara}]{hirama2004}
Hirama, M., Tokosumi, T., Ishida, T., \& ichi Aihara, J. 2004, Chemical
  Physics, 305, 307

\bibitem[{Hoekstra {et~al.}(1991)Hoekstra, de~Heer, \&
  Morgenstern}]{hoekstra1991}
Hoekstra, R., de~Heer, F.~J., \& Morgenstern, R. 1991, Journal of Physics B:
  Atomic, Molecular and Optical Physics, 24, 4025

\bibitem[{{Hornek{\ae}r} {et~al.}(2006){Hornek{\ae}r}, {Rauls}, {Xu}, {{\v
  S}ljivan{\v c}anin}, {Otero}, {Stensgaard}, {L{\ae}gsgaard}, {Hammer}, \&
  {Besenbacher}}]{hornekaer2006}
{Hornek{\ae}r}, L., {Rauls}, E., {Xu}, W., {et~al.} 2006, Physical Review
  Letters, 97, 186102

\bibitem[{{Le Page} {et~al.}(2009){Le Page}, {Snow}, \&
  {Bierbaum}}]{lepage2009}
{Le Page}, V., {Snow}, T.~P., \& {Bierbaum}, V.~M. 2009, \apj, 704, 274

\bibitem[Li \& Draine (2012)]{li2012}
Li, A. \& Draine, B.~T. 2012, ArXiv e-prints, 1210.6558

\bibitem[{Martinazzo \& Tantardini(2006)}]{martinazzo2006chem}
Martinazzo, R. \& Tantardini, G.~F. 2006, Journal of Chemical Physics, 124

\bibitem[{{Maziarz}(2005)}]{maziarz2005}
{Maziarz}, E. 2005, Canadian Journal of Chemistry - Revue Canadienne de Chimie,
  85, 1871

\bibitem[{{Mennella}(2006)}]{mennella2006}
{Mennella}, V. 2006, \apjl, 647, L49

\bibitem[{{Mennella} {et~al.}(2012){Mennella}, {Hornek{\ae}r}, {Thrower}, \&
  {Accolla}}]{mennella2012}
{Mennella}, V., {Hornek{\ae}r}, L., {Thrower}, J., \& {Accolla}, M. 2012,
  \apjl, 745, L2

\bibitem[{Morisset {et~al.}(2003)Morisset, Aguillon, Sizun, \&
  Sidis}]{morisset2003}
Morisset, S., Aguillon, F., Sizun, M., \& Sidis, V. 2003, Chemical Physics
  Letters, 378, 615

\bibitem[{Morisset {et~al.}(2004b)Morisset, Aguillon, Sizun, \&
  Sidis}]{morisset2004b}
Morisset, S., Aguillon, F., Sizun, M., \& Sidis, V. 2004b, The Journal of
  Physical Chemistry A, 108, 8571

\bibitem[{{Oort} \& {van de Hulst}(1946)}]{oort1946}
{Oort}, J.~H. \& {van de Hulst}, H.~C. 1946, \bain, 10, 187

\bibitem[{{Perry} \& {Price}(2003)}]{perry2003}
{Perry}, J.~S.~A. \& {Price}, S.~D. 2003, \apss, 285, 769

\bibitem[{{Pirronello} {et~al.}(2000){Pirronello}, {Biham}, {Manic{\'o}},
  {Roser}, \& {Vidali}}]{pirronello2000}
{Pirronello}, V., {Biham}, O., {Manic{\'o}}, G., {Roser}, J.~E., \& {Vidali},
  G. 2000, in Molecular Hydrogen in Space, ed. {F.~Combes \& G.~Pineau Des
  Forets}, 71--+

\bibitem[{{Pirronello} {et~al.}(1999){Pirronello}, {Liu}, {Roser}, \&
  {Vidali}}]{pirronello1999}
{Pirronello}, V., {Liu}, C., {Roser}, J.~E., \& {Vidali}, G. 1999, \aap, 344,
  681

\bibitem[{{Pirronello} {et~al.}(1997){Pirronello}, {Liu}, {Shen}, \&
  {Vidali}}]{pirronello1997a}
{Pirronello}, V., {Liu}, C., {Shen}, L., \& {Vidali}, G. 1997, \apjl, 475, L69+

\bibitem[{{Rauls} \& {Hornek{\ae}r}(2008)}]{rauls2008}
{Rauls}, E. \& {Hornek{\ae}r}, L. 2008, \apj, 679, 531

\bibitem[{{Rougeau} {et~al.}(2006){Rougeau}, {Teillet-Billy}, \&
  {Sidis}}]{rougeau2006}
{Rougeau}, N., {Teillet-Billy}, D., \& {Sidis}, V. 2006, Chemical Physics
  Letters, 431, 135

\bibitem[{{Sha} \& {Jackson}(2002)}]{sha2002}
{Sha}, X. \& {Jackson}, B. 2002, Surface Science, 496, 318

\bibitem[{{Shah} \& {Gilbody}(1978)}]{shah1978}
{Shah}, M.~B. \& {Gilbody}, H.~B. 1978, Journal of Physics B Atomic Molecular
  Physics, 11, 121

\bibitem[{{Snow} {et~al.}(1998){Snow}, {Le Page}, {Keheyan}, \&
  {Bierbaum}}]{snow1998}
{Snow}, T.~P., {Le Page}, V., {Keheyan}, Y., \& {Bierbaum}, V.~M. 1998, \nat,
  391, 259

\bibitem[{{Spaans} \& {Meijerink}(2005)}]{spaans2005}
{Spaans}, M. \& {Meijerink}, R. 2005, \apss, 295, 239

\bibitem[{{Spaans} \& {Silk}(2000)}]{spaans2000}
{Spaans}, M. \& {Silk}, J. 2000, \apj, 538, 115

\bibitem[{{Thrower} {et~al.}(2012){Thrower}, {J{\o}rgensen}, {Friis},
  {Baouche}, {Mennella}, {Luntz}, {Andersen}, {Hammer}, \&
  {Hornek{\ae}r}}]{thrower2012}
{Thrower}, J.~D., {J{\o}rgensen}, B., {Friis}, E.~E., {et~al.} 2012, \apj, 752,
  3

\bibitem[{{Tielens} {et~al.}(1993){Tielens}, {Meixner}, {van der Werf},
  {Bregman}, {Tauber}, {Stutzki}, \& {Rank}}]{tielens1993}
{Tielens}, A.~G.~G.~M., {Meixner}, M.~M., {van der Werf}, P.~P., {et~al.} 1993,
  Science, 262, 86

\bibitem[{{Weingartner} \& {Draine}(2001)}]{Weingartner2001}
{Weingartner}, J.~C. \& {Draine}, B.~T. 2001, \apj, 548, 296

\bibitem[{{Wolfire} {et~al.}(2008){Wolfire}, {Tielens}, {Hollenbach}, \&
  {Kaufman}}]{wolfire2008}
{Wolfire}, M.~G., {Tielens}, A.~G.~G.~M., {Hollenbach}, D., \& {Kaufman}, M.~J.
  2008, \apj, 680, 384

\bibitem[{{Zecho} {et~al.}(2002){Zecho}, {Guttler}, {Sha}, {Jackson}, \&
  {Kuppers}}]{zecho2002}
{Zecho}, T., {Guttler}, A., {Sha}, X., {Jackson}, B., \& {Kuppers}, J. 2002,
  \jcp, 117, 8486

\end{thebibliography}
\end{document}